\documentclass[letterpaper]{article}
\usepackage{aaai}
\usepackage{times}
\usepackage{helvet}
\usepackage{courier}
\usepackage{amsmath,amssymb,amsfonts}
\usepackage{algorithmic}
\usepackage{graphicx}
\usepackage{textcomp}
\usepackage{xcolor}
\usepackage{soul}
\usepackage{url}
\usepackage[utf8]{inputenc}
\usepackage{amsmath}
\usepackage{amsthm}
\usepackage{booktabs}
\usepackage{algorithm}
\usepackage{algorithmic}
 \usepackage{xcolor}

\usepackage{scalerel}
\def\pitslab{\kern0em\raise-.01em\hbox{\scalerel*{\includegraphics{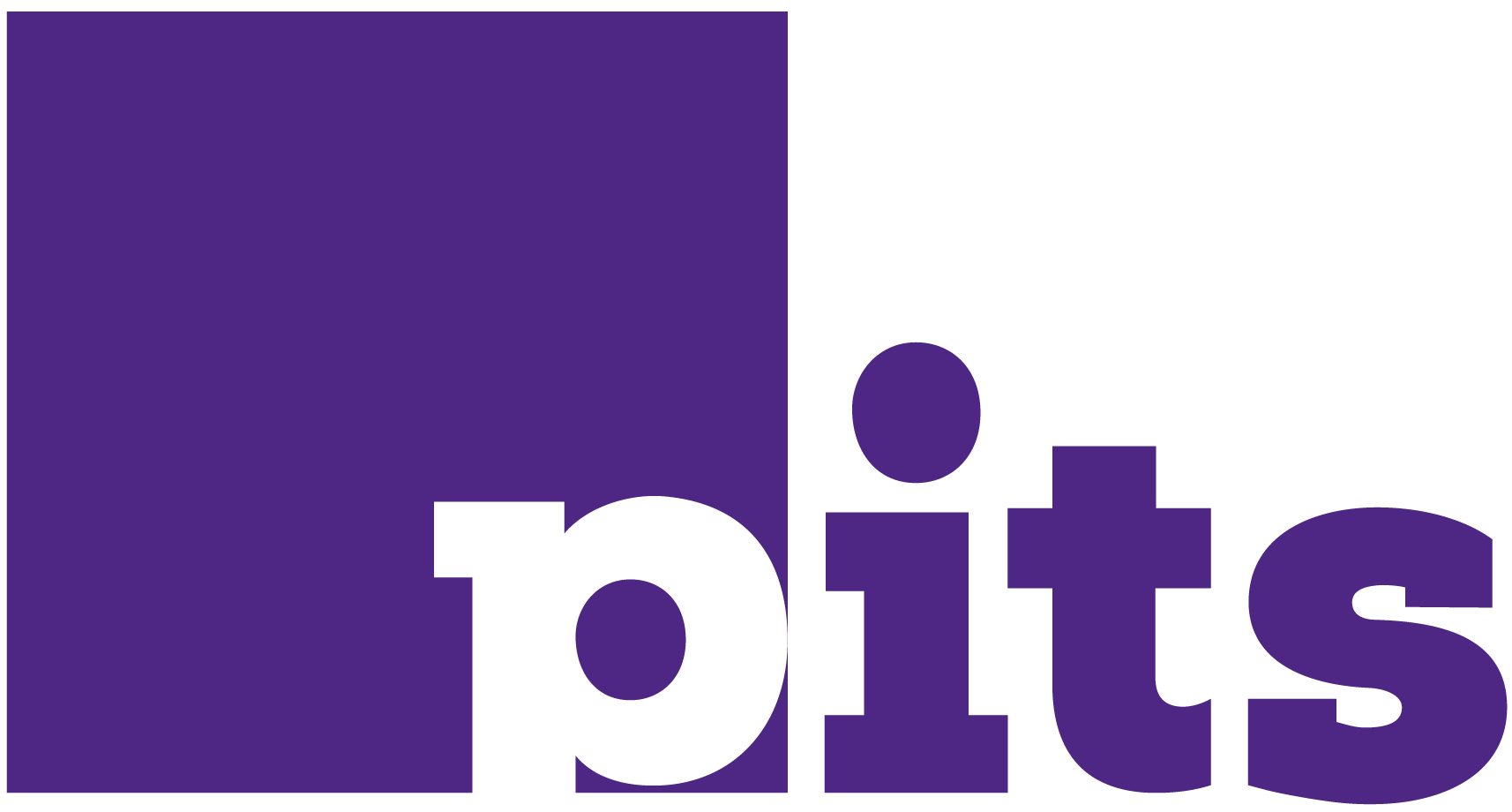}}{X}}}
 
\begin{document}
\pagestyle{plain}
\title{Implicit Crowdsourcing for Identifying Abusive Behavior in Online Social Networks}
\author{Abiola Osho \and Ethan Tucker \and George Amariucai\\
Dept. of Computer Science\\
Kansas State University\\ 
\{aaarise, ectucker, amariucai\}@ksu.edu \\}

\maketitle
\begin{abstract}
\begin{quote}
The increased use of online social networks for the dissemination of information comes with the misuse of the internet for cyberbullying, cybercrime, spam, vandalism, amongst other things. To proactively identify abuse in the networks, we propose a model to identify abusive posts by crowdsourcing. The crowdsourcing part of the detection mechanism is implemented implicitly, by simply observing the natural interaction between users encountering the messages. We explore the node-to-node spread of information on Twitter and propose a model that predicts the abuse level (abusive, hate, spam, normal) associated with the tweet by observing the attributes of the message, along with those of the users interacting with it. We demonstrate that the difference in users' interactions with abusive posts can be leveraged in identifying posts of varying abuse levels.
\end{quote}
\end{abstract}

\section{Introduction}
The use of online social networks (OSNs) like Twitter for the dissemination of information around the world continues to increase in today's society. Due to their wide reach, oversimplified conversations, anonymity, and ability to provide quick blasts of information, online social networks have also become an avenue for various cyberbullying and harassment. The common exchange and banters in social networking platforms can sometimes become explosive, resulting in uncontrolled expression of hate, harassment, and aggression without the risk of reciprocal physical injury or personal danger. In recent times, there have been reports of mental and psychological health issues directly linked to abuse, harassment, and cyberbullying \cite{singleton2016online}\cite{landstedt2014bullying}\cite{hebert2016child}\cite{goebert2011impact}. To curtail these abusive behaviors, there is a need to quickly identify posts intended for abuse, and have them expunged from the network to reduce users' exposure.

Not all messages shared with abusive intent are written crudely. A message intended as abuse might be concealed in sarcasm, emotions, and sentiments even if it appears otherwise. We hypothesize that there is a difference in the diffusion of abusive posts in OSNs, a difference that can be leveraged by using a crowdsourcing approach to predict the abusive label associated with these posts. We believe that some users are more likely than others to create, share, and/or react to abusive posts. These users will serve as discriminators in the detection model to sieve out outliers who do not contribute much to the detection task. We adopt an implicit crowdsourcing model by simply observing users' interaction with posts of varying abuse levels to ensure that the user's posting and reaction behavior is as natural as it can be. In this study, we introduce an automated model for predicting the abuse level associated with a tweet predicated on the interaction between users encountering the messages. We describe two types of posts, \textit{normal} and \textit{abusive}. A post is said to be \textit{abusive}, if and only if the content or context associated aligns with the intent for cyber abuse. Seeing as feature design and selection strongly impact a machine learning model's accuracy much more than the model used \cite{hall1999correlation}, we train a Bayesian Logistic Regression model by incorporating user, message, and propagation features to estimate the node-to-node influence dynamics to the propagation of abusive posts. We describe two tasks in identifying abusive behavior online and adopt supervised machine learning models in addressing them. 
\begin{enumerate}
    \item We investigate abusive behavior prediction by exploring abuse propagation founded on microscopic-level information spread. By observing the spreading behavior of posts of varying abuse levels in online social networks, we propose a model that predicts the abuse level associated with a tweet by observing the latent attributes of the message, along with those of the users, and their reactions over the network.
    \item We evaluate the role of user and message features in detecting the abuse level of a post, by measuring the contributions of individual users and their posts to the spread of abusive posts in OSNs.
\end{enumerate}


Previous crowdsourcing-based approaches in abuse detection in social networks focus on conversation or account annotation for abuse detection. To the best of our knowledge, this is the first research that explores crowdsourcing as an automated tool for identifying abusive behavior in online social networks. We classify users based on the types of posts they generally react to: (i) reacts to only \textit{normal} posts, (ii) reacts to only \textit{abusive} posts and (iii) reacts to a mix of \textit{normal} and \textit{normal} posts. Users in class i and ii are good discriminators for both abuse level detection and feature identification, while users in class iii do not serve as good discriminators in the prediction model. The contributions of this paper are as follows:
\begin{itemize}
    \item We introduce a new paradigm to abusive behavior identification that applies implicit crowdsourcing for predicting the abuse level of a post without user annotation.
    \item We present a model to predict the abuse level associated with a tweet using the propagation pattern and achieve accuracy up to 20\% higher than models that do not incorporate propagation patterns.
    \item We demonstrate the abilities of the crowdsourced model by presenting a ranking of features relevant to abusive post propagation.
\end{itemize}

The paper is organized as follows: Section \ref{prevwork} reviews the related work on detecting abusive behavior, and features that aid information spread in social networks. Section \ref{method} describes the prediction tasks, proposed model, and dataset. Section \ref{result} presents experimental results and observations, and finally, Section \ref{summary} gives conclusions and insights into possible future works.


\section{Related Works}\label{prevwork}
Research on detecting abusive behavior online has explored numerous techniques including machine learning, deep learning, crowdsourcing, amongst others. In this section, we provide a brief overview of some of the studies and methods relating to abusive behavior identification and detection. We provide an additional review of feature selection for information propagation to emphasize the role of feature selection in detection tasks.


\subsection{Feature Selection for Information Propagation}
\cite{guille} introduced a variant of the AsIC model called the T-BAsIC framework that assigns a fixed value for a real time-dependent function for each link, without fixing the diffusion probability. The model relies on three different dimensions to compute the diffusion probability: social, semantic, and time. The model was designed to predict the daily volume of tweets for a topic and variations in popularity of topics over time. They proceeded by identifying 2 types of users: (1) transmitters that pass along information and (2) stiflers that become dead-ends for information travel, with stiflers growing with time for a given topic.  In \cite{ferrara2016predicting}, the authors leverage a mixture of metadata, network, and temporal features in detecting users spreading extremist ideology and predict content adopters and interaction reciprocity in social media. They adopted logistic regression and random forests learning models with 52 features observed from Twitter data of over 25,000 accounts labeled as supportive of the Islamic State. 
Given the temporal relevance of tweets, \cite{spasojevic2015post} propose finding the best times for a user to post on social networks in order to maximize the probability of audience response. They hypothesize that the probability that an audience member reacts to a message depends on factors such as his daily and weekly behavior patterns, his location and timezone, and the volume of other messages competing for his attention.

\subsection{Models for Abusive Behavior Detection}
As OSNs become interesting targets for spammers and malicious users, \cite{verma2014techniques} reviewed literature to identify features used for detecting spam and malicious users. They pointed out that spam detection algorithms commonly explore features categorized as user-based, content-based, and hybrid (combining user and content-based features. \citeauthor{badjatiya2017deep} applied several deep learning models with pre-trained word embedding over a dataset of 16k labeled tweets to classify tweets as racist, sexist, or neither. The best results from their experiments were derived by training with Long Short Term Memory networks (LSTMs) and embedding with a Gradient Boosted Decision Trees (GBDT) \cite{badjatiya2017deep}. The work of \cite{nobata2016abusive} describes a supervised learning system for detecting abusive language in online comments. Token unigrams and bigrams, and character n-grams were extracted from a dataset of 2 million Yahoo finance and news article comments. They adopted a classifier based on linguistic and syntactic features to achieve an F-score of 0.795 for Finance comments and 0.817 for News comments. To classify Twitter users as aggressors, bullies, or spammers,  \cite{chatzakou2017mean} developed a classifier based on random forests to identify aggressive and bullying accounts. The model used a combination of user, text, and network features for its identification task. \citeauthor{almaatouq2016if} presented an analysis of suspended spam accounts on Twitter. Using Gaussian Mixture Model, the authors discovered that there are two primary categories of spammers on Twitter with distinct behavior. They hypothesized that the first group mainly consists of fraudulent accounts, while the second is made up of legitimate accounts that have been compromised.

\subsection{Crowdsourcing Techniques for Offensive Behavior Identification}
CrowdFlower is a popular tool among researchers for labeling data for research requiring labeled data as ground truth. The researchers in \cite{burnap2015cyber} used CrowdFlower to label 2000 tweets by having annotators answer the question "Is this text offensive or antagonistic in terms of race, ethnicity, or religion?". They used the labeled data in a machine learning classifier for identifying hateful and antagonistic content on Twitter. \citeauthor{founta2018large} used CrowdFlower to annotate a large collection of tweets with a set of abuse-related labels. Their research covers different forms of abusive behavior in order to identify a robust and consistent set of labels - abusive, hateful, normal, and spam - to characterize abuse-related tweets. To distinguish between hate speech and everyday usage of potentially offensive language in tweets, \cite{ICWSM1715665} presented an automated model to classify tweets as hate speech, offensive language, or neither. They used labeled data crowdsourced using CrowdFlower and adopted a logistic regression model with L2 regularization to identify hate speech.

On examining the mislabeled hateful tweets in the work of \cite{ICWSM1715665}, they observed that some were possibly incorrectly labeled in the crowdsourcing step, and some contained few of the terms commonly associated with hate speech. Tweets with less common slurs were also frequently mislabeled. One of the challenges to crowdsourcing is to ensure workers provide objective and truthful reporting. A lot of sites rely on posters to crowdsource the identification of abusive content because it is impossible for moderators to identify all abusive content. The research of \cite{ghosh2011moderates} presented an algorithm where users rate content based on a set of ratings and the users also get rated based on the probability that they will correctly label a contribution. To account for the trustworthiness of crowdsourced content, \cite{crowdsourceincentive} proposed a bidding and incentive mechanism for mobile crowdsourcing. To ensure trustworthy submissions, the authors applied Evolutionary Game Theory to ensure that the best strategy for workers was to submit trustworthy data. Each worker is assigned a reputation score, which begins at a maximum but is decreased if a worker submits untrustworthy data. It is also increased if the worker submits trustworthy data. Different tasks on the platform have different reputation thresholds, which workers must exceed to work on the task. This makes reporting trustworthy data the most stable strategy for workers.

In crowdsourcing tasks, especially when backed by incentives, participants may introduce an implicit bias in the data brought about by the ``presence" of an observer, leading to a change in behavior \cite{jccl19} or opinion and causing them to provide feedback that they feel is expected or sense what the ``community" rewards, and comply. By contrast, our proposed crowdsourcing mechanism 
aims to observe users in the wild, making it less subject to the bias introduced by conscious detection.



\section{Model and Method}\label{method}
We propose a framework that given a tweet will predict the abuse level by observing the user interaction with the tweet --  we leverage ``the wisdom of the crowd" as it is often used in a crowdsourcing approach to assigning a label to the post. The proposed model differs from current crowdsourcing techniques in that it makes an inference from a supervised learning task and does not require a human annotator. Since this is a supervised learning task, the model requires labeled data and makes use of manually annotated tweets for learning and inference. To each user, we associate a total of 16 features, including 3 network and 13 interaction attributes; and to each message, we associate 11 attributes. We then train a Bayesian logistic regression model based on the prediction task. 

\subsection{Data Description}
In this study, we make use of the ICWSM 2020 task 2 dataset made publicly available by \cite{founta2018large}. The dataset contains 100k annotated tweets associated with inappropriate speech labeled as abusive and hateful speech, as well as normal interactions and spam. Because the model we propose makes use of attributes of the user for inference, we used a tweet Hydrator - an Electron-based desktop application for hydrating Twitter ID datasets \cite{hydrator}. The Hydrator helps us turn tweet IDs back into JSON, retrieving information contained in the Tweet and User Objects.

Since tweets get deleted from time to time, by the user or Twitter, some tweets were no longer available through the Twitter API, as such, we had a reduced number of tweets after hydration. The dataset contained over 69k tweets with 56k unique users. Tweets labeled normal made up 62\% of the dataset, abusive tweets accounted for 20\%, spam tweets constituted 14\%, while hateful tweets formed 4\% of the data. 

We recreate the Twitter followership graph for the available dataset by associating an edge between two users if there is a follower-followee relationship between them. Based on the assumption that users will interact with their friends' messages uniquely, we assign the diffusion label as a function of the reaction observed per message and show that this microscopic-level information spread based on the latent message and user interaction attributes is sufficient to give insight to the abuse level of a message.


\subsection{Task 1: Implicit Crowdsourcing for Predicting the Abuse Level of a Tweet}\label{task1}
\begin{figure}[h]
    \includegraphics[width=\linewidth]{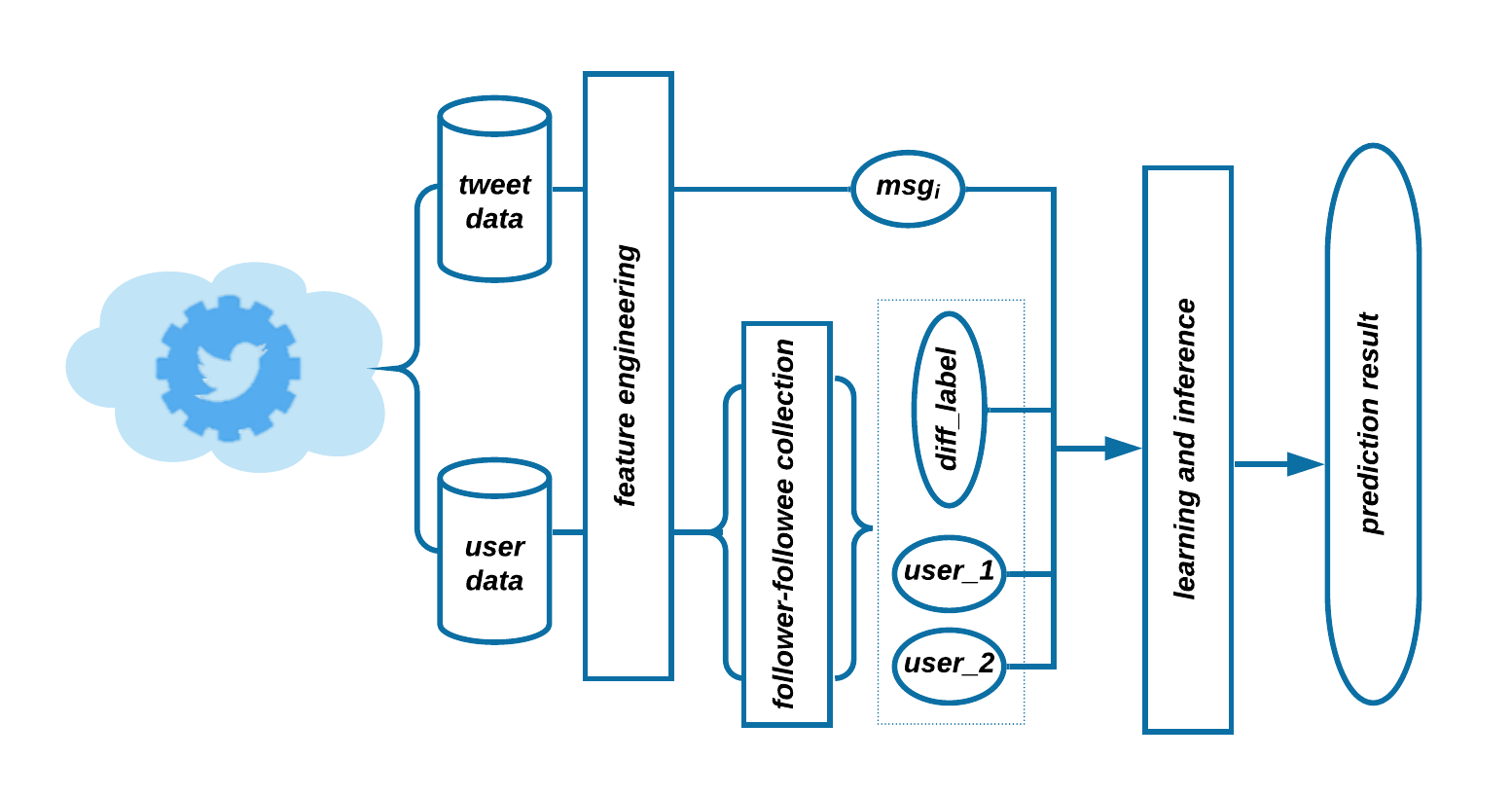}
    \caption{Model for implicit crowdsourcing for predicting abuse level of a tweet}
    \label{fig:methodology}
  \end{figure}

First, we demonstrate the differences in the propagation of \textit{Abusive} and \textit{Normal} posts, we perform a node-to-node analysis between a pair of users, the spreader and receiver, examining each user's posting behavior, and their interactions to predict the receiver's reaction. Here, we aim to show that our model performs well in an established environment, to compare with previous models for propagation prediction. This task is valuable to strengthening our hypothesis that the propagation behavior is a significant attribute to predicting the abuse level of a message based on how users in OSN interact with posts of varying veracity. We believe that a tweet that is abusive or hate speech will stir up reaction from many users in the network, causing it to propagate farther than a normal or spam post will. Then, we train a model that predicts the abusive label associated with a message. We extract the features described in Section \ref{features}, and additionally include the diffusion property as an independent variable during the training phase, see Figure \ref{fig:methodology} for a conceptual illustration of the model. 

More specifically, an edge is said to be \emph{diffused} if and only if the destination user (in Twitter terms: \textit{follower}) has reacted (reply, retweet, quote, like) to the friend's (\textit{followee's}) post. We examine how users on Twitter relate with posts of their friends by building classifiers to distinguish user interactions based on the abusive label associated with the message. For a message $m$, where $m \in \{1,\ldots,M\}$, spreading over a network with $n$ interactions, we train a model that predicts the abusive label associated with the message based on the diffusion behavior observed along each one of the $n$ links along which the message propagates. The predicted output is the majority abusive label observed across the $n$ interactions. For instance, if an \textit{abusive} message is spread over 5 interactions and the model predicts the post to be \textit{abusive} 3 out of 5 times, we accept the output to be \textit{abusive} and evaluate the model over its correct classification of $M$ messages in the test collection.

By comparison, the non-crowdsourced model relies solely on the features of the message and those of the original creator of the message in making a decision on the truth-status of the tweet.



\subsection{Task 2: Estimating Features Contributing to Abusive Tweet Propagation}\label{task2}

We perform a supervised learning task where we train the model using the attributes from a pair of nodes with an established followership relationship and label the interaction between them as either diffused or not diffused. The attributes learned are said to be representative of users' network, interaction,  participation, role, and importance in the spread of information to other nodes in the network. As previously stated, these attributes are learned over four different time intervals. After learning these features, we fit a regression function that maps the learned user attributes to the likelihood of diffusion between the nodes. In this task, we describe two sub-tasks:
\begin{enumerate}
    \item We build a persona for the user to evaluate the user's tendency to post or react to posts of different abuse levels.
    \item We explore the node-to-node relationship between users and seek to identify features that cause abusive posts to propagate.
\end{enumerate}

For each user in the dataset, we assign an abusive, hate, spam, or normal score computed as a ratio of their post that is labeled as such. For each label $i$, where $i \in \{abusive, hate, spam, normal\}$

\begin{equation}\label{score}
score_i = \frac{count_i}{N}
\end{equation}
where N is the total number of tweets the user has in the collection.

\subsubsection{Task 2.1: Evaluating a user's tendency to post or reactive to abusive posts}

For the purpose of this data challenge, we assume a user's total tweets to be limited to the data in the collection. However, for a more robust prediction task, it is important that the score estimated in eqn (\ref{score}) is estimated over the tweets shared on the user's timeline. For each user, we create an online persona by combining the user features, message-based features (over all of the user's messages) and the estimated abusive scores. We perform regression analysis on this behavioral pattern and then train a Random Forest classifier to rank the features that directly impact the probability that a user will post or react to a message that is labeled abusive or hateful. Currently, we limit the prediction task to focus on estimating a user's probability to post or react to abusive and/or hateful posts as these kinds of behavior are not as widely studied as spam. 

\subsubsection{Task 2.2: Identifying features for abusive post propagation}
We perform a supervised learning task where we train the model using the attributes from a pair of nodes with an established followership relationship and label the interaction between them as either diffused or not diffused. The attributes learned are said to be representative of the user's profile, messages,  network, interaction,  participation, role, and importance in the spread of information to other nodes in the network. After learning these features, we fit a regression function that maps the learned user attributes to the likelihood of diffusion between the nodes. Then, we present a ranking of the features that contribute to the likelihood of abusive and hate tweets diffusing over the network.

\subsection{Feature Description}\label{features}
We suggest 3 categories of features: message, interaction, network, and train a random forest classifier to rank the features in order of importance. The features highlighted here are combined with the diffusion label as part of the input variables in the bayesian logistic regression model used in Task 1. 
 
\subsubsection{Network-based features}
In microblogs such as Twitter, a \emph{friend} is someone a user follows, and a user can see all of his friend's posts. In like manner, a follower is someone that follows and has direct access to all of a user's posts. We consider three features of the user's network: \textit{followers count}, \textit{friends count}, which have been extensively studied by \cite{castillo2011information}\cite{liang2015rumor}\cite{yang2012automatic}, and \textit{followers to friends ratio}, which was used in \cite{wu2015false} to establish opinion leaders. These attributes are important because a user's friends impact the kind and volume of messages that end up in his timeline and the higher the number of followers, the farther the possibility of reach. This is also reflected in policies by OSNs like Twitter and Instagram who attach value to the follower count, where users become verified once they cross a certain threshold, even if the account holder is not a celebrity or public figure. Table \ref{network} describes the network features used in the model.

\begin{table}[H]
\footnotesize
\centering
\begin{tabular}{p{1in}p{2in}}   
\toprule
Feature  &  Description \\
\midrule
followers count & higher count depict higher reach\\
friends count & \# of accounts user follows\\
followers-friend & ratio to show influence in the network \\
\bottomrule
\end{tabular}
\caption{Network-based features}
\label{network}
\end{table}

\subsubsection{Interaction-based Features}
We identify specific attributes of the user's online persona and posting behavior as a contributor to the user exhibiting abusive behavior online. The assumption is that both the follower and followee contribute equally to the diffusion of a post, and an aggregate of network and message attributes tilt the reaction decision. Table \ref{interaction} describes the 13 interaction attributes being considered.

\begin{table}[H]
\footnotesize
\centering
\begin{tabular}{p{0.9in}p{2.1in}}
\toprule
Feature  &  Description \\
\midrule
directed tweets & ratio of tweets directed at someone\\
dialogue & active interaction from user 1 to 2\\
retweet-to-tweet & ratio of user's tweets with retweet\\
tweet wit hashtag & ratio of user's tweets that contain hashtags\\
tweets with url & ratio of user's posts with URL\\
tweets with media & ratio of user's posts with media\\
avg favorite-tweet & ratio of posts that get favorited\\
avg tweets/day & shows how active the user is\\
has url	& does user's profile have a URL\\
has description & does user's profile have description\\
is verified	& is the account verified\\
status count &	volume of tweets over account's lifetime\\
account age	& \# of days since account was created\\
\bottomrule
\end{tabular}
\caption{Interaction-based features}
\label{interaction}
\end{table}

\subsubsection{Message-based Features}
We perform content analysis by adopting the sentiment analysis framework provided by TextBlob \cite{loria2018textblob} to assign sentiment score to the message. Table \ref{message} describe the message attributes adopted in our model. 

\begin{table}[H]
\footnotesize
\centering
\begin{tabular}{p{0.9in}p{2.1in}} 
\toprule
Feature  &  Description \\
\midrule
quoted status & has post been quoted \\
is rt& has post been retweet\\
rt count& \# of retweets\\	
rt status & is post a retweet \\
favorited count& \# of favorites\\
has hashtag& does post contain hashtags\\
has url	& does post contain URL\\
has mentions& does post mention someone using ``@"\\
has media& does post contain media\\
avg tweet length& length of tweet / 280 (max length)\\
sentiment score & polarity of tweet\\
\bottomrule
\end{tabular}
\caption{Message-based Features}
\label{message}
\end{table}

\subsection{Evaluation Metrics}\label{eval}
The prediction capabilities of the learned model are tested based on its abilities to predict if there is diffusion across an edge given the learned model. We use standard classification evaluation metrics: precision, recall and, F score, to assess the efficiency of our model.

\textbf{Precision} describes the ratio of instances correctly classified as ``diffused'' to the total classified as ``diffused'', and is estimated as:

\begin{equation}
     Precision = \frac{TP}{TP + FP}
\end{equation}

\textbf{Recall} is the ratio of instances correctly classified as ``diffused'' to the total number of instances that ``diffused", and is estimated as:
\begin{equation}
     Recall = \frac{TP}{TP + FN},
\end{equation}

\noindent
Where $TP$ (true positives) is the number of instances correctly classified as ``diffused", $FP$ (false positives) is the number of instances incorrectly classified as ``diffused", and $FN$ (false negatives) is the number of the instances incorrectly classified as `` not diffused".

The \textbf{F score} is the harmonic mean of the precision and recall. It is computed as 
\begin{equation}
      F score = 2 \times \frac{Precision \times Recall }{Precision + Recall}
\end{equation}


\section{Results}\label{result}

\subsection{Predicting Abuse Level using Implicit Crowdsourcing}\label{result_task1}

By simply observing the reaction generated between users in the network, we train a model that learns to distinguish the interaction-reaction relationship. The model's ability to effectively distinguish the uniqueness of this relationship over messages of different abuse levels is useful in detecting the abuse label associated with a message by observing the reaction and in turn, propagation of the message over the network. As previously stated, the objective of this task is to show that collating the implicitly sourced diffusion behavior between users is useful for detecting the abusive behavior of a post. This implicit crowdsourcing approach is important in real-world situations where there is a need for the system to passively interact with the network.


\begin{table*}[ht!]
\centering
\begin{tabular}{lcccccc}
\hline
Abuse level & CRO - Precision & CRO - Recall  & CRO - F1  & nonCRO - Precision  & nonCRO - Recall & nonCRO - F1 \\
\hline
Abusive & 0.85& 0.82& \textbf{0.83} & 0.65& 0.71& \textbf{0.68}\\
Hate & 0.82 & 0.89 & \textbf{0.85} & 0.61 & 0.63& \textbf{0.62}\\
Spam & 0.90 & 0.92  & \textbf{0.91} & 0.67 & 0.69& \textbf{0.68}\\
Normal & 0.95 & 0.90 & \textbf{0.92} & 0.80 & 0.85 & \textbf{0.82}\\
Abusive+Hate & 0.94 & 0.97 & \textbf{0.95} & 0.74 & 0.79& \textbf{0.76}\\
\hline
\end{tabular}
\scriptsize\caption{Model performance in abuse level of a post}
\label{task1_result}
\end{table*}

We carry out prediction tasks to detect the abuse levels associated with a tweet and we included an additional task to predict if a post is offensive (abusive OR hate). In Table \ref{task1_result}, we show the performance of the model using the evaluation metrics described in Section \ref{eval}, with the dataset split in a 60-30-10 train-validation-test ratio. We present the precision, recall, and F1 scores for the crowdsourced model (CRO - *) and differentiate it from the non-crowdsourced (nonCRO - *) model. From the results, we observed that prediction tasks using the crowdsourced model performed considerably better recording over 20\% improvement than the non-crowdsourced model. The prediction result for posts labeled as \textit{normal} is unsurprising because the model had more data to learn from than the other labels. For a model to implicitly predict the abuse level associated with a tweet, there is a need to learn from the user's prior interactions with the network. We further argue that a user's likelihood to create or react to an offensive post will influence his/her future interactions with similar posts. One can also argue that increased exposure of a user to offensive posts in his/her network will increase his chances of posting the same.


\subsection{Features for Abusive Behavior Propagation}\label{result_task2}
Here, we model the user's participation in the spread of abusive posts on Twitter and use the knowledge to measure the contribution of individual user in the creation and spread of abusive posts in OSNs.

\subsubsection{Features impacting user's propensity for abusive posts}
On Twitter, a user can show his/her interest in a topic by contributing to the topic through the creation of posts, retweets, replies, quotes, etc. By contributing to a given topic, users give little hints into their interests, possibly patterns to their behavior and expected reactions. We group tweets labeled as abusive and hate together as abusive posts. Even though the data is heavily skewed towards normal posts, result from the experiment shows that the sentiment around the topic plays a major role in whether a user will post something abusive about it. Following closely to sentiment is the abusive score, this is unsurprising because a user with a high abusive score will most likely keep posting abusive tweets. As expected, the normal score ranks side by side the abusive score as they complement each other. Ranked next to that are the user favorite and average tweet per day. We observed that users with more friends than followers are more likely to exhibit abusive behaviors. 

These features are only descriptive of the user's own tendencies towards abusive posts. It is important to note that identifiable events in the network can also contribute to a user's disposition to share abusive posts at a particular point in time. The presence of media (such as memes, emojis, and sometimes images with text) poses a challenge to this task as some of these media might contain offensive content that the model is unable to interpret. Due to the fluidity of the Twitter interaction, language, and user interests, we believe this task will perform better as a semi-supervised learning task where the model learns to adapt to the dynamic nature of the Twitter network. 

\subsubsection{Features impacting abusive post propagation}
Establishing a difference in the diffusion prediction models for \textit{abusive} and \textit{normal} posts is amply dependent on showing that there exists a difference between these types of messages and the attributes that steer user reactions. In previous tasks, we have shown that the detection models differ from one abusive label to another, here, we show that the messages propagate differently by providing evidence that the attributes contributing to diffusion differ between \textit{abusive} and \textit{normal} posts. Please recall that \textit{abusive} posts are described as tweets in the data associated with abusive and hate labels.

To further validate our assumption that there is a difference associated with the message, interaction and diffusion patterns of \textit{abusive} and \textit{normal} posts, we use random forest classifiers to provide the top-10 features, see Table \ref{task2_rank} that aid in the propagation of \textit{abusive} and \textit{normal} messages. In this task, we model the diffusion of posts from one user to another and observe the reaction of the receiving user. We learn a function that maps what features of the source and destination users cause a reaction or otherwise. We measure the model's ability to correctly predict a user's reaction based on the learned function.

\begin{table}[ht!]
\footnotesize
\centering
\begin{tabular}{|c|l|l|} 
\hline
Rank & Abusive  & Normal \\
\hline
1 & dest friends count & MSG is RT\\
2 & dialogue &  MSG has mentions\\
3 & MSG has mentions & MSG favorited count\\              
4 & dest tweet with hashtag & src tweets with URL\\
5 & src retweet-to-tweet & dest retweet-to-tweet \\    
6 & src status count  & MSG sentiment score\\         
7 & src followers count & src directed tweet\\
8 & MSG sentiment score & dialogue\\       
9 & src followers-friends & src avg favorite-tweet\\
10 &  MSG has hashtag & dest follower-friends\\   
\hline
\end{tabular}
\scriptsize\caption{Top 10 features for predicting propagation of \textit{abusive} and \textit{normal} posts selected using Random Forest classifiers}
\label{task2_rank}
\end{table}

From the ranked features, we see that the sentiment score and established dialogue is deemed important in the diffusion of posts between two users. The network features are ranked as important to the spread of abusive posts. One thing to note here is that a single user can act as both the source or destination node in the network, depending on his role at a particular point in time. Additionally, the presence of hashtag(s) in a tweet has an effect on its likelihood to get a reaction.


\section{Conclusion}\label{summary}


In this study, we hypothesize that there is a difference in the spreading behavior of posts with varying abuse levels. We presented a model based on Bayesian logistic regression to predict the abuse level of a message by simply crowdsourcing the interaction and propagation behaviors of similar messages. The crowdsourcing detection model integrates information diffusion by using the diffusion label (``diffused" or ``not diffused") associated with the node-to-node interaction between a pair of users. Results from our experiment show an improvement of about 20\% over models that are non-crowdsourced.

One way to extend this work will be to assign an abusive score for a user based on the emotions his/her posts incite in the network and the responses the messages get i.e., a user will be deemed more abusive if he incites abusive and/or hateful responses within the network.
~{\Large \pitslab}

\section*{Acknowledgments}
\addcontentsline{toc}{section}{Acknowledgment}
This work was supported in part by the U.S. National Science Foundation under Grants No. 1527579 and 1619201.


\bibliographystyle{aaai}
\bibliography{main}
\end{document}